\documentclass[a4paper]{article}

\usepackage{INTERSPEECH2020}
\usepackage{mathabx}
\usepackage{hyperref}

\title{Neural Spatio-Temporal Beamformer for Target Speech Separation}
\name{Yong Xu$^1$, Meng Yu$^1$, Shi-Xiong Zhang$^1$,  Lianwu Chen$^2$, Chao Weng$^1$, Jianming Liu$^1$, Dong Yu$^1$}
\address{
  $^1$Tencent AI lab, Bellevue, WA, USA ~~~~~~~~ $^2$Tencent AI lab, Shenzhen, China}
\email{\{lucayongxu, raymondmyu, auszhang, lianwuchen, cweng, jianmingliu, dyu\}@tencent.com}

\begin{document}

\maketitle
\begin{abstract}
    Purely neural network (NN) based speech separation and enhancement methods, although can achieve good objective scores, inevitably cause nonlinear speech distortions that are harmful for the automatic speech recognition (ASR). On the other hand, the minimum variance distortionless response (MVDR) beamformer with NN-predicted masks, although can significantly reduce speech distortions, has limited noise reduction capability. In this paper, we propose a multi-tap MVDR beamformer with complex-valued masks for speech separation and enhancement. Compared to the state-of-the-art NN-mask based MVDR beamformer, the multi-tap MVDR beamformer exploits the inter-frame correlation in addition to the inter-microphone correlation that is already utilized in prior arts. Further improvements include the replacement of the real-valued masks with the complex-valued masks and the joint training of the complex-mask NN. The evaluation on our multi-modal multi-channel target speech separation and enhancement platform demonstrates that our proposed multi-tap MVDR beamformer improves both the ASR accuracy and the perceptual speech quality against prior arts.

\end{abstract}

\section{Introduction}



The deep learning based speech enhancement  \cite{wang2013towards,xu2014regression,lu2013speech} and speech separation \cite{hershey2016deep,yu2017permutation,chen2017deep} methods have attracted lots of research attention since the renaissance of the neural network. However, the purely neural network based front-end approaches inevitably cause nonlinear speech distortions \cite{du2014robust}. The speech distortion can degrade the performance of the speech recognition system \cite{du2014robust}, even for the commercial general-purpose ASR engine which is already robust enough to the background noise. The refinement \cite{du2014robust} or joint training \cite{wang2016joint,gao2015joint,qian2018single} on the enhanced speech can make the front-end output and the back-end acoustic model match better. Nevertheless, these approaches cannot explicitly reduce the speech distortion. Furthermore, the joint training with the commercial general-purpose ASR engine is usually not feasible either because the training data is too large and noisy or because the ASR engine is third-party.

For example, the fully-convolutional time-domain audio separation network (Conv-TasNet) \cite{luo2019conv} has shown significant improvement in the speech separation task. We further proposed several audio-visual \cite{wu2019time} or multi-channel \cite{tan2019audio, bahmaninezhad2019comprehensive,lorry2020} speech separation techniques based on the Conv-TasNet. Although these models can obtain substantial gain according to the objective measures \cite{luo2019conv, wu2019time, bahmaninezhad2019comprehensive}, they cause some nonlinear distortions in the separated speech because such distortion is not considered for attenuation in the model. 
%

On the other hand, the minimum variance distortionless response (MVDR) beamformer \cite{souden2009optimal}, as its name suggests, explicitly requires distortionless filtering on the target direction \cite{benesty2008microphone} and thus has significantly less speech distortions in the separated speech. Recently, MVDR have been improved by exploiting better covariance matrix computation through NN estimated ideal ratio masks (IRMs) \cite{heymann2016neural, erdogan2016improved, xu2019joint, aswin2020icassp,wang2018mask}. Although NN-mask based MVDR  \cite{boeddeker2018exploring,xiao2017time} can achieve better ASR accuracy than purely NN-based approaches due to less distortions, the residual noise level of the enhanced speech is high.

\begin{figure*}[htb]
	\begin{minipage}[b]{1.0\linewidth}
		\centering
		\centerline{\includegraphics[width=0.90\textwidth]{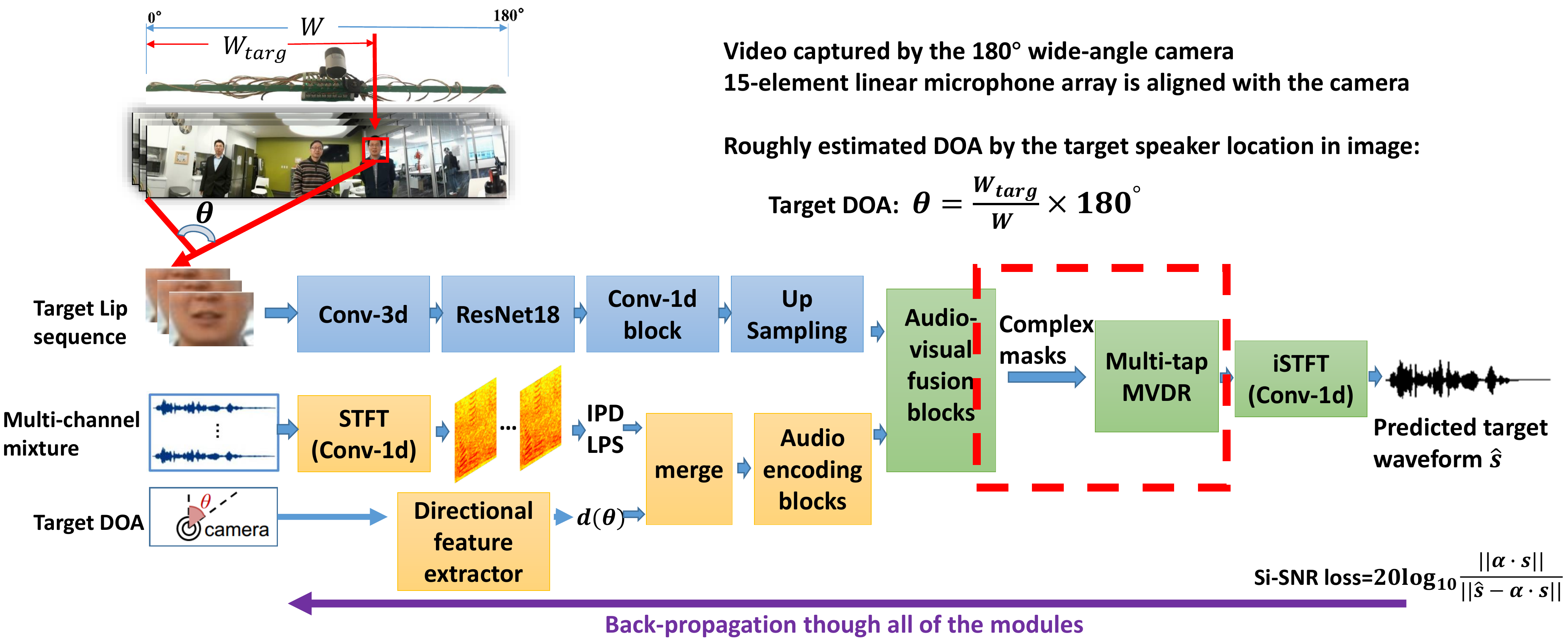}}
	\end{minipage}
	\caption{Joint training of the multi-tap MVDR with complex-valued masks. The \textbf{complex masking} and \textbf{multi-tap MVDR},  highlighted in the dashed rectangle, which are the focus of this paper. $\bf{\alpha}=\hat{\bf{s}}^\text{T}\bf{s}/\bf{s}^\text{T}\bf{s}$ is a scaling factor in the time-domain Si-SNR loss.} 
	\label{fig:overview_system}
\end{figure*}

In this work, we propose a neural spatio-temporal beamforming approach, named multi-tap MVDR beamformer with complex-valued masks, for speech separation and enhancement to simultaneously obtain high ASR accuracy and PESQ score. The multi-tap MVDR for the multi-channel scenario is inspired by the multi-frame MVDR on the single channel \cite{schasse2014estimation, huang2011multi, tammen2019dnn, benesty2011speech, fischer2017sensitivity}. Similar to the MVDR, multi-tap MVDR enforces distortionless at the target direction. Different from the MVDR and multi-frame MVDR, which utilize the inter-microphone correlation and inter-frame correlation, respectively, the multi-tap MVDR exploits both correlations and thus has higher potential. Benesty et al. \cite{benesty2011speech} proposed a similar idea for the multi-channel speech enhancement from the signal processing perspective. Our proposed approach differentiates with theirs in that ours is NN-mask based. Additional novelties in our approach include the replacement of the real-valued masks \cite{hershey2016deep, erdogan2016improved, lorry2020,tan2019audio} with the complex-valued masks (CMs), and the joint training of the CMs in the multi-tap MVDR framework. We evaluated our proposed approach on our multi-modal multi-channel target speech separation platform \cite{tan2019audio,lorry2020} by replacing the speech separation component shown in Fig. \ref{fig:overview_system}. Our experiments indicate that the multi-tap MVDR beamformer with CMs improves both the ASR accuracy and the perceptual speech quality against prior arts.

The rest of the paper is organized as follows. In Section \ref{sec:cm_mtmvdr}, we describe our proposed multi-tap MVDR beamformer with complex-valued masks. In Section \ref{sec:exp} we present the baseline system and the experimental setup. The results are given in Section \ref{sec:result}. We conclude the paper in Section \ref{sec:conclude}.

\section{Neural Spatio-Temporal Beamformer: Multi-tap MVDR with Complex Mask}\label{sec:cm_mtmvdr}

\subsection{Spatial filtering: MVDR beamformer}
MVDR is a widely used beamformer for ASR \cite{heymann2016neural}. It minimizes the power of the noise (interfering speech + additive noise) while ensuring that the signal at the desired direction is not distorted. Mathematically, this can be formulated as,
\begin{equation}
	\bf{w}_{MVDR}=\underset{\bf{w}}{\arg\min}\bf{w}^{\sf H}\bf{\Phi_{NN}}\bf{w}  ~~~s.t.~~~  \bf{w}^{\sf H}\bf{v}=1
	\label{mvdr_constraint}
\end{equation}
Where ${\bf{\Phi_{NN}}}(f) \in \mathbb{C}^{M\times M}$ is the covariance matrix of noise $\bf{N}$ at frequency bin $f$ and $\bf{v}$ is the target steering vector. $M$ is the number of the microphone. The constraint $\bf{w}^{\sf H}\bf{v}=1$ is important to guarantee that the target source is distortionless. There are several solution variants for this optimization problem \cite{souden2009optimal,habets2013two}. 
The solution based on the reference channel selection \cite{subramanian2019investigation,habets2013two, souden2009optimal} is
\begin{equation}
{\bf{w}_{MVDR}}(f)=\frac{{\bf{\Phi^{-1}_{NN}}}(f){\bf{\Phi_{SS}}}(f)}{\text{Trace}({{\bf\Phi^{-1}_{NN}}(f)}{\bf{\Phi_{SS}}}(f))}\textbf{u}~~~, ~~~~~{\bf{w}}(f)\in \mathbb{C}^{M}
\label{mvdr_solution}
\end{equation}



where $\bf{u}$ is the one-hot vector representing a reference microphone channel and $\bf{\Phi_{SS}}$ represents the covariance matrix of the target speech. The key step for the beamforming is to estimate the two covariance matrices, namely $\bf{\Phi_{NN}}$ and $\bf{\Phi_{SS}}$. For the traditional signal processing based techniques, the noise frames and speech frames are tracked to update $\bf{\Phi_{NN}}$ and $\bf{\Phi_{SS}}$ in a recursive way. Research indicates that better results can be achieved using the mask-based covariance matrix estimation method with neural networks \cite{heymann2016neural}.

\subsection{Neural spatial filtering: Mask based MVDR}

The idea behind the mask based approach for covariance matrix estimation is that we may more accurately estimate the target speech, and thus the covariance matrix, given a NN-based mask estimator (will discuss in Sec. \ref{sec:vis_encoder}). The most commonly used mask for the mask-based beamforming \cite{heymann2016neural} is ideal ratio mask (IRM) \cite{wang2014training} or sigmoid mask. In this work, we extend to use ReLU-Mask and linear uncompressed complex-valued mask for the covariance matrix calculation. The ReLU-mask (a.k.a. STFT magnitude mask) \cite{wang2014training} is defined as,
\begin{equation} \label{eq:relu_mask}
\text{ReLU-Mask}(t,f)=\frac{|S(t,f)|}{|Y(t,f)|}
\end{equation}
Where $|S|$ and $|Y|$ represents the target speech magnitude and noisy speech magnitude, respectively. The range of ReLU-Mask lies in [0,+$\infty$]. Note that no value clipping is needed in our implementation, which is different from the FFT-MASK in \cite{wang2014training} where the value was clipped into [0,10]. This is because our scale-invariant source-to-noise ratio (Si-SNR) \cite{luo2019conv} loss function (shown in Fig. \ref{fig:overview_system}) is optimized on the recovered time-domain waveform rather than on the mask itself. 

Given the real-valued mask (RM) (as the output of a sigmoid or ReLU function) defined on the magnitude, the covariance matrix $\bf{{\Phi}}_{\textbf{SS}}$ of the beamformer can be computed as
\begin{equation}
{\bf{{\Phi}}}_{\textbf{SS}}(f)=\frac{\sum_{t=1}^{T}{\text{RM}}^2_\textbf{S}(t,f)\textbf{Y}(t,f)\textbf{Y}^{\sf H}(t,f)}{\sum_{t=1}^{T}{\text{RM}}^2_\textbf{S}(t,f)}
\end{equation}
Where $T$ is the chunk size. 
We argue that better covariance matrix estimation can be achieved with the complex-valued mask (CM) for speech separation in this work and ASR \cite{xu2019joint, yu2020audio}. 
The $\text{CM}(t,f)$ was first proposed in \cite{williamson2015complex} as,
\begin{equation}
{S}={S}_r+j{S}_i=(\text{CM}_r+j\text{CM}_i)*(Y_r+jY_i)=\text{CM}*{Y}
\label{eq:cm}
\end{equation}
where $r$ and $i$ denote the real part and the imaginary part of the complex spectrum, respectively. The theoretical range of CM lies in [$-\infty$,$+\infty$]. In \cite{williamson2015complex}, the CM was compressed into [-10,10] since their model was trained to estimate the CM itself. 
In our implementation, however, value compression is not necessary and can be harmful. We implicitly estimate the CM with a linear activation function and then multiply it with the complex spectrum of the mixture to obtain the estimated clean speech. The Si-SNR loss \cite{luo2019conv} function is optimized on the reconstructed time-domain waveform rather than on CM itself. 
With CM, $\bf{{\Phi}}_{\textbf{SS}}$ can be rewritten as
\begin{align}
{\bf{{\Phi}}}_{\textbf{SS}}(f)&=\frac{\sum_{t=1}^{T}\hat{\textbf{S}}(t,f)\hat{\textbf{S}}^{\sf H}(t,f)}{\sum_{t=1}^{T}{\text{CM}}^{\sf H}_\textbf{S}(t,f)\text{CM}_\textbf{S}(t,f)} \\
&=\frac{\sum_{t=1}^{T}({\text{CM}}_\textbf{S}(t,f)\textbf{Y}(t,f))({\text{CM}}_\textbf{S}(t,f)\textbf{Y}(t,f))^{\sf H}}{\sum_{t=1}^{T}{\text{CM}}^{\sf H}_\textbf{S}(t,f)\text{CM}_\textbf{S}(t,f)}
\end{align}
where $M_S$ and $\text{CM}_S$ are shared across channels. The mask normalization in the denominator is the key to success since the weighted mask is to attend on the most related frames to calculate $\bf{{\Phi}}$. ${\bf{{\Phi}}}_{\textbf{NN}}(f)$ can be computed in the similar way.
According to the MVDR solution Eq. (\ref{mvdr_solution}), the beamformed speech of the target speaker can be estimated by,
\begin{equation}
\hat{\bf{S}}(t,f)={{\bf{w}^{\sf H}}(f){\bf{Y}}}(t,f)
\label{mvdr}
\end{equation}
\subsection{Neural spatio-temporal filtering: CM based Multi-tap MVDR}
Although MVDR can improve the ASR performance, it keeps the speech distortion low at the cost of high residual noise \cite{habets2013two}. 
Inspired by the single channel multi-frame MVDR \cite{huang2011multi, tammen2019dnn, fischer2017sensitivity} which utilizes the inter-frame correlation, we propose a multi-tap MVDR for the multi-channel neural beamforming to achieve distortionless speech and low residual noise simultaneously. We define the $L$-tap representation of the mixture speech as $\bar{\textbf{Y}}(t,f)=[\textbf{Y}^\mathsf{T}(t,f),\textbf{Y}^\mathsf{T}(t-1,f),...,\textbf{Y}^\mathsf{T}(t-L+1,f)]^\mathsf{T} \in \mathbb{C}^{M\times L}$. The corresponding $\bar{\textbf{S}}$, $\bar{\textbf{N}}$, $\widebar{\textbf{CM}}$  can be defined in the same way. Then we can calculate the extended $L$-tap target speech covariance matrix ${\bf{{\Phi}}}_{{\bar{\textbf{S}}\bar{\textbf{S}}}}(f) \in \mathbb{C}^{ML \times ML}$ as
\begin{equation}
{\bf{{\Phi}}}_{{\bar{\textbf{S}}\bar{\textbf{S}}}}(f)=\frac{\sum_{t=1}^{T}({\widebar{\text{CM}}}_\textbf{S}(t,f)\bar{\textbf{Y}}(t,f))({\widebar{\text{CM}}}_\textbf{S}(t,f)\bar{\textbf{Y}}(t,f))^{\sf H}}{\sum_{t=1}^{T}{\widebar{\text{CM}}}^{\sf H}_\textbf{S}(t,f)\widebar{\text{CM}}_\textbf{S}(t,f)}
\end{equation}
Benesty et al. \cite{benesty2011speech} proposed the multi-channel speech enhancement filter. However, our approach is different from theirs in that we are using complex-valued masks estimated by neural networks to compute the covariance matrix. Similar to Eq. (\ref{mvdr_solution}), the multi-tap MVDR solution is
\begin{equation}
{\bar{\textbf{w}}}(f)=\frac{{{\bf{\Phi^{-1}_{\bar{N}\bar{N}}}}(f)}{\bf{\Phi_{\bar{S}\bar{S}}}}(f)}{\text{Trace}({{\bf{\Phi^{-1}_{\bar{N}\bar{N}}}}(f)}{{\bf{\Phi_{\bar{S}\bar{S}}}}(f)})}{\bar{\textbf{u}}}~~~, ~~~~~{\bar{\bf{w}}}(f)\in \mathbb{C}^{M\times L}
\label{mtmvdr_solution}
\end{equation}
where $\bar{\bf{u}}$ is an expanded one-hot vector of $\bf{u}$ with padding zeros in the tail. Note that the multi-tap MVDR follows the optimization process of MVDR in Eq. (\ref{mvdr_constraint}) for the multi-channel scenario. Hence, it is different from the multi-frame MVDR (MFMVDR) \cite{huang2011multi,tammen2019dnn} defined on the single channel. The enhanced speech of the multi-tap MVDR can be obtained as,
\begin{equation}
\hat{\textbf{S}}(t,f)=\bar{\textbf{w}}^{\sf H}(f)\bar{\textbf{Y}}(t,f)
\label{mtmvdr}
\end{equation}
The beamformed spectrum is converted to the time-domain waveform via iSTFT. Finally, the Si-SNR loss \cite{luo2019conv} calculated on the waveform is back-propagated through all of the modules (including the multi-tap MVDR module and the networks) as shown in Fig. \ref{fig:overview_system}. Different from the weighted prediction error (WPE) \cite{kinoshita2017neural} for dereverberation, multi-tap MVDR utilizes the correlation of nearest frames (mainly the early reflection area) and aims only at recovering the reverberant clean speech. However, WPE keeps away from the early reflection area to avoid hurting the dry clean speech for the dereverberation \cite{kinoshita2017neural}.

In summary, the solution Eq. (\ref{eq:cm}) is a complex-valued masking on the single channel. MVDR provides a solution Eq. (\ref{mvdr}) of complex masking on multiple channels, and our proposed multi-tap MVDR (Eq. (\ref{mtmvdr})) conducts spatio-temporal filtering across frames and channels.

\section{Experimental Setup and Baselines} \label{sec:exp}

We evaluate our proposed methods on our multi-modal multi-channel target speech separation platform \cite{tan2019audio,lorry2020}. The audio-visual structure is shown in Fig. \ref{fig:overview_system} and briefly overviewed below.

\subsection{Multi-modal multi-channel mask estimator baseline}\label{sec:vis_encoder}
As shown in Fig. \ref{fig:overview_system}, we use the direction of arrival (DOA) of the target speaker and the speaker-dependent lip sequence for informing the dilated convolutional neural networks (CNNs) to extract the target speech from the multi-talker mixture.

\textbf{Video encoder}: The captured video can provide two important speaker-dependent information, lip movement sequence and the DOA of the target speaker (denoted as $\theta$ in Fig. \ref{fig:overview_system}).
The lip movement has been proven effective for the speech separation in \cite{ephrat2018looking, afouras2018conversation, morrone2019face, wu2019time, lorry2020}.
In this work, we utilize the mouth region RGB pixels to represent the target speaker's lip feature. As shown in Fig. \ref{fig:overview_system}, a 3-D residual network \cite{afouras2018conversation,chung2016lip,stafylakis2017combining} is adopted to extract the target speech related lip movement embeddings.

\textbf{Audio encoder}: The audio input includes the speaker-independent features (e.g., log-power spectra (LPS) and interaural phase difference (IPD) \cite{tan2019audio}) and speaker-dependent feature (e.g., directional feature $d(\theta)$ \cite{chen2018multi,lorry2020}).
%
As shown in Fig. \ref{fig:overview_system}, the 15-element non-uniform linear microphone array \cite{tan2019audio} is co-located with the $180^{\circ}$ wide-angle camera. The location of the target speaker's face in the whole camera view can provide a rough DOA estimation of the target speaker. Chen et al \cite{chen2018multi} proposed a location guided directional feature (DF) $d(\theta)$ to extract the target speech from the specific DOA. DF aims at calculating the cosine similarity between the target steering vector $v({\theta})$ and IPDs \cite{chen2018multi}. 
The LPS, IPDs and DF are merged and fed into a bunch of dilated 1D-CNNs. The details can be found in our previous work \cite{tan2019audio,lorry2020}.

Then the concatenated lip embeddings and audio embeddings \cite{tan2019audio} are used to predict the sigmoid mask (i.e., IRM) or the ReLU-mask (Eq. (\ref{eq:relu_mask})) used in our previous work \cite{tan2019audio}, or the complex-valued mask (as Eq. (\ref{eq:cm})) proposed in this study. 

\subsection{Dataset and experimental setup}
\begin{table*}[htbp]
	\centering
	\caption{PESQ and WER results of some dilated CNN baselines and proposed jointly trained multi-tap MVDR system.}
	\begin{tabular}{l|cccc|ccc|c|c}
		\hline
		Systems/Metrics & \multicolumn{7}{c|}{PESQ $\in[-0.5,4.5]$}                              & PESQ   & WER (\%) \\
		\hline
		& \multicolumn{4}{c|}{ Angle between target \& others}    & \multicolumn{3}{c|}{ \# of overlapped speakers} &         &  \\
		\hline
		& 0-15{$^\circ$}    & 15-45{$^\circ$}    & 45-90{$^\circ$}    & 90-180{$^\circ$}   & 1 SPK & 2 SPK & 3 SPK   & Ave   & Ave \\
		\hline
		Reverberant Clean (reference) & 4.50  & 4.50  & 4.50  & 4.50  & 4.50  & 4.50  & 4.50  & 4.50   & 6.97\% \\
		Mixture (interfering speech + noise) & 1.88  & 1.88  & 1.98  & 2.03  & 3.55  & 2.02  & 1.77  & 2.16  & 51.30\% \\
		ReLU mask (Audio only) on Channel 0 (i) & 2.50  & 2.68  & 2.88  & 2.86  & 3.88  & 2.81  & 2.50  & 2.87   & 17.89\% \\
		ReLU mask (Lip only) on Channel 0 (ii) & 2.44  & 2.52  & 2.74  & 2.68  & 3.86  & 2.76  & 2.34  & 2.76   & 23.25\% \\
		ReLU mask (Baseline) on Channel 0 (iii) & 2.56  & 2.74  & 2.93  & 2.89  & 3.88 & 2.85  & 2.56  & 2.92   & 17.44\% \\
		\textbf{Complex mask (CM) on Channel 0} (iv) & 2.64  & 2.84  & 3.00  & 3.00  & \textbf{3.89} & 2.94  & 2.66  & 3.00  & 16.90\% \\
		\hline
		Sigmoid mask MVDR joint train (JT) (v) & 2.27  & 2.59  & 2.82  & 2.73  & 3.67  & 2.67  & 2.37  & 2.73    & 15.11\% \\
		ReLU mask MVDR JT (vi) & 2.52  & 2.74  & 2.94  & 2.85  & 3.68  & 2.86  & 2.54  & 2.88   & 12.61\% \\
		CM MVDR JT (vii) & 2.55  & 2.77  & 2.97  & 2.89  & 3.73  & 2.89  & 2.57  & 2.91  & 11.84\% \\
		\textbf{Prop. CM multi-tap MVDR JT} (viii) & \textbf{2.70} & \textbf{3.00} & \textbf{3.20} & \textbf{3.13} & 3.83  & \textbf{3.10} & \textbf{2.76} & \textbf{3.10}  & \textbf{9.96\%} \\
		\hline
	\end{tabular}%

	\label{tab:pesq}%
\end{table*}%

The mandarin audio-visual corpus \cite{tan2019audio} used for experiments is collected from Youtube. We use SNR estimation tool and face detection tool to filter out low SNR ($\le$ 17dB) and multi-face videos \cite{tan2019audio}, resulting in 205500 clean video segments with single face (about 200 hours) over 1500 speakers. The sampling rate for audio and video are 16 kHz and 25 fps respectively. 512-point of STFT is used to extract audio features along 32ms Hann window with 50\% overlap. A mouth region (size=112x112x3) detection program \cite{lorry2020} is run on the target speaker's video to capture the the lip movements.

The new and larger multi-talker multi-channel far-field dataset are simulated in the similar way with our previous work \cite{tan2019audio,lorry2020}.
The simulated dataset contains 190000, 15000 and 500 multi-channel mixtures for training, validation and testing. The speakers in the test set are unseen in the training set. The transcript of the speech for the ASR evaluation is manually labeled by human in this work.
The multi-channel signals are generated by convolving speech with RIRs simulated by image-source method \cite{habets2006room}.
The signal-to-interference ratio (SIR) is ranging from -6 to 6 dB. Also, noise with 18-30 dB SNR is added to all the multi-channel mixtures \cite{tan2019audio}. 
A commercial general-purpose mandarin speech recognition Yitu API \cite{yitu} (uncorrelated to this work) is used to test the ASR performance.

The multi-modal network is trained in a chunk-wise mode with chunk size 4 seconds, using Adam optimizer with early stopping. Initial learning rate is set to 1e-3. The $L$-tap in the multi-tap MVDR is set to 3 empirically. Pytorch 1.1.0 was used.

\section{Results and Discussions}\label{sec:result}

The PESQ and ASR word error rate (WER) results are shown in Table \ref{tab:pesq} to compare among purely network-based systems and several jointly trained MVDR systems. Note that we only conduct speech separation and denoising without dereverberation in this work. Our systems work well on different scenarios, e.g., different angles between the target speaker and other speakers, various number of overlapped speakers. The scenarios, e.g., small angles ($\le$45$^\circ$) or more overlapped speakers, are a bit more challenging.

\begin{figure}[htb]
	\begin{minipage}[b]{1.0\linewidth}
		\centering
		\centerline{\includegraphics[width=0.9\textwidth]{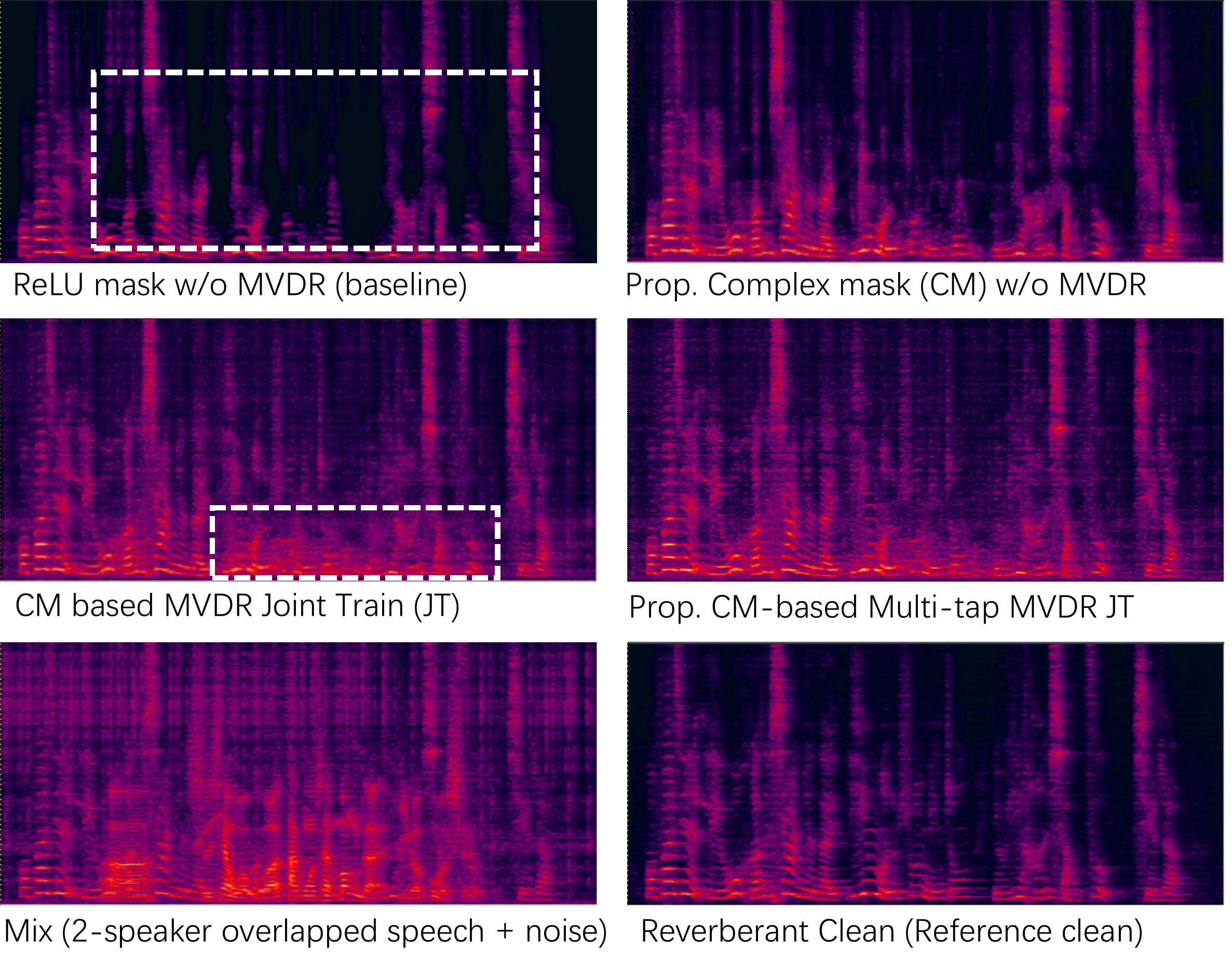}}
	\end{minipage}
	\caption{Separated spectrogram  \href{https://yongxuustc.github.io/mtmvdr}{demos} of different systems.} 
	\label{fig:spec}
\end{figure}

\textbf{Real-valued mask VS CM}: The linear uncompressed complex mask (CM) based system (iv) achieves higher PESQ (3.00 vs 2.92) and lower WER (16.90\% vs 17.44\% ) compared to the ReLU mask baseline (iii). The difference between the ReLU mask and CM is also shown in Fig. \ref{fig:spec}. There are some spectral ``black holes'' distortion in the enhanced spectrogram of the ReLU mask baseline (iii).
The problem is more severe in the sigmoid mask according to our observations. 
This type of nonlinear spectral distortion is harmful to speech recognition. However, the CM can reduce the distortion and recover the phase simultaneously.
Better mask can also help to estimate more accurate covariance matrix in the MVDR beamformer.

\textbf{Mask-based MVDR}: Although the CM-based network (iv) can reduce the distortion and achieve 3.00 PESQ on average, the ASR performance of 16.9\% WER does not match the gain in PESQ. This phenomenon is widely observed in purely neural network based speech enhancement front-ends \cite{du2014robust} because the non-linear distortion in the enhanced speech is not ASR friendly. Even for the commercial general-purpose ASR engine, non-distortion is more important than no residual noise, considering that the ASR engine is already robust to the mild level noise (but may not robust to the non-linear distortion). With the distortionless constraint in the MVDR, the beamformed speech can achieve much lower WER. For example, the jointly trained CM-based MVDR (vii) can reduce the WER from 16.9\% to 11.84\% when compared to the CM-based network w/o MVDR (iv). CM is superior to other real-valued masks (ReLU mask or sigmoid mask) in estimating the target speech and noise covariance matrix. Nonetheless, MVDR beamformer obtains this distortionless advantage by sacrificing the strength of residual noise reduction \cite{habets2013two}, e.g., the jointly trained CM-based MVDR (vii) only achieves 2.91 PESQ on average and is lower than purely network-based system (iv) with 3.00 PESQ.
 
\textbf{CM-based multi-tap MVDR}: The proposed jointly trained complex mask based multi-tap MVDR (viii) can get the best average PESQ, i.e., 3.10 and lowest WER, i.e., 9.96\%, surpassing the best purely network-based system (iv).
Compared to the common MVDR (vii), the multi-tap MVDR (viii) can achieve about 0.2 PESQ improvement on the 2/3-speaker cases since the multi-tap MVDR can utilize the inter-frame correlation and reduce the uncorrelated noise. The difference is also shown in Fig. \ref{fig:spec} where the proposed multi-tap MVDR can reduce more residual noise while ensuring the distortionless constraint. More demos (including real-world testing demos) can be found at our website: \href{https://yongxuustc.github.io/mtmvdr}{https://yongxuustc.github.io/mtmvdr}. 

\textbf{Directional feature VS lip feature}: As introduced in Sec. \ref{sec:vis_encoder}, two speaker dependent features are used in this work, namely lip features and the DF ($d(\theta)$). Although multi-modality evaluation is not the focus of this work, we compare the audio only (using $d(\theta)$ w/o lip) and the lip only (using lip w/o $d(\theta)$) setup for the ablation study. The audio only system (i) is better than the lip only system (ii) (WER 17.89\% vs 23.25\%). It indicates that the DF ($d(\theta)$) is more distinct than lip feature. But when the two modalities are concatenated together to form the system (iii), slightly better performance can be achieved with WER 17.44\%. More analysis about the multi-modality fusion can be found in our previous work \cite{tan2019audio,lorry2020}.  




\section{Conclusions and Future Work}\label{sec:conclude}
In this work, we proposed the multi-tap MVDR with complex-valued masks (CMs). We demonstrated that CM can achieve less distortion and better ASR performance for the purely neural network based systems, and can better estimate the covariance matrix in the mask-based beamformer, than the real-valued masks. With the proposed CM based multi-tap MVDR, we obtain both the best ASR performance and PESQ among all systems. Compared to the purely neural network baseline using ReLU-mask, multi-tap MVDR can significantly reduce the WER from 17.44 \% to 9.96\% and improve the PESQ from 2.92 to 3.10 on average. 
We want to emphasize that the results achieved with multi-tap MVDR indicates that using filter-based instead of mask-based models for speech separation is promising. We will further extend the spatio-temporal filtering to spatio-temporal-frequency filtering and conduct separation and dereverberation in an integrated framework.


\newpage

\bibliographystyle{IEEEtran}

\bibliography{mybib}

\begin{thebibliography}{10}
\providecommand{\url}[1]{#1}
\csname url@samestyle\endcsname
\providecommand{\newblock}{\relax}
\providecommand{\bibinfo}[2]{#2}
\providecommand{\BIBentrySTDinterwordspacing}{\spaceskip=0pt\relax}
\providecommand{\BIBentryALTinterwordstretchfactor}{4}
\providecommand{\BIBentryALTinterwordspacing}{\spaceskip=\fontdimen2\font plus
\BIBentryALTinterwordstretchfactor\fontdimen3\font minus
  \fontdimen4\font\relax}
\providecommand{\BIBforeignlanguage}[2]{{%
\expandafter\ifx\csname l@#1\endcsname\relax
\typeout{** WARNING: IEEEtran.bst: No hyphenation pattern has been}%
\typeout{** loaded for the language `#1'. Using the pattern for}%
\typeout{** the default language instead.}%
\else
\language=\csname l@#1\endcsname
\fi
#2}}
\providecommand{\BIBdecl}{\relax}
\BIBdecl

\bibitem{wang2013towards}
Y.~Wang and D.~Wang, ``Towards scaling up classification-based speech
  separation,'' \emph{IEEE Transactions on Audio, Speech, and Language
  Processing}, vol.~21, no.~7, pp. 1381--1390, 2013.

\bibitem{xu2014regression}
Y.~Xu, J.~Du, L.-R. Dai, and C.-H. Lee, ``A regression approach to speech
  enhancement based on deep neural networks,'' \emph{IEEE/ACM Transactions on
  Audio, Speech, and Language Processing}, vol.~23, no.~1, pp. 7--19, 2014.

\bibitem{lu2013speech}
X.~Lu, Y.~Tsao, S.~Matsuda, and C.~Hori, ``Speech enhancement based on deep
  denoising autoencoder.'' in \emph{Interspeech}, 2013, pp. 436--440.

\bibitem{hershey2016deep}
J.~R. Hershey, Z.~Chen, J.~Le~Roux, and S.~Watanabe, ``Deep clustering:
  Discriminative embeddings for segmentation and separation,'' in
  \emph{ICASSP}, 2016.

\bibitem{yu2017permutation}
D.~Yu, M.~Kolb{\ae}k, Z.-H. Tan, and J.~Jensen, ``Permutation invariant
  training of deep models for speaker-independent multi-talker speech
  separation,'' in \emph{ICASSP}, 2017.

\bibitem{chen2017deep}
Z.~Chen, Y.~Luo, and N.~Mesgarani, ``Deep attractor network for
  single-microphone speaker separation,'' in \emph{ICASSP}, 2017.

\bibitem{du2014robust}
J.~Du, Q.~Wang, and et~al., ``Robust speech recognition with speech enhanced
  deep neural networks,'' in \emph{Interspeech}, 2014.

\bibitem{wang2016joint}
Z.-Q. Wang and D.~Wang, ``A joint training framework for robust automatic
  speech recognition,'' \emph{IEEE/ACM Transactions on Audio, Speech, and
  Language Processing}, vol.~24, no.~4, pp. 796--806, 2016.

\bibitem{gao2015joint}
T.~Gao, J.~Du, L.-R. Dai, and C.-H. Lee, ``Joint training of front-end and
  back-end deep neural networks for robust speech recognition,'' in
  \emph{ICASSP}, 2015, pp. 4375--4379.

\bibitem{qian2018single}
Y.~Qian, X.~Chang, and D.~Yu, ``Single-channel multi-talker speech recognition
  with permutation invariant training,'' \emph{Speech Communication}, vol. 104,
  pp. 1--11, 2018.

\bibitem{luo2019conv}
Y.~Luo and N.~Mesgarani, ``Conv-tasnet: Surpassing ideal time--frequency
  magnitude masking for speech separation,'' \emph{IEEE/ACM transactions on
  audio, speech, and language processing}, vol.~27, no.~8, pp. 1256--1266,
  2019.

\bibitem{wu2019time}
J.~Wu, Y.~Xu, S.-X. Zhang, L.-W. Chen, M.~Yu, L.~Xie, and D.~Yu, ``Time domain
  audio visual speech separation,'' \emph{ASRU}, 2019.

\bibitem{tan2019audio}
K.~Tan, Y.~Xu, S.-X. Zhang, M.~Yu, and D.~Yu, ``Audio-visual speech separation
  and dereverberation with a two-stage multimodal network,'' \emph{IEEE Journal
  of Selected Topics in Signal Processing}, 2020.

\bibitem{bahmaninezhad2019comprehensive}
F.~Bahmaninezhad, J.~Wu, R.~Gu, S.-X. Zhang, Y.~Xu, M.~Yu, and D.~Yu, ``A
  comprehensive study of speech separation: spectrogram vs waveform
  separation,'' \emph{Interspeech}, 2019.

\bibitem{lorry2020}
R.~Gu, S.-X. Zhang, Y.~Xu, L.~Chen, Y.~Zou, and D.~Yu, ``Multi-modal
  multi-channel target speech separation,'' \emph{IEEE Journal of Selected
  Topics in Signal Processing}, 2020.

\bibitem{souden2009optimal}
M.~Souden, J.~Benesty, and S.~Affes, ``On optimal frequency-domain multichannel
  linear filtering for noise reduction,'' \emph{IEEE Transactions on audio,
  speech, and language processing}, vol.~18, no.~2, pp. 260--276, 2009.

\bibitem{benesty2008microphone}
J.~Benesty, J.~Chen, and Y.~Huang, \emph{Microphone array signal
  processing}.\hskip 1em plus 0.5em minus 0.4em\relax Springer Science \&
  Business Media, 2008.

\bibitem{heymann2016neural}
J.~Heymann, L.~Drude, and R.~Haeb-Umbach, ``Neural network based spectral mask
  estimation for acoustic beamforming,'' in \emph{ICASSP}, 2016.

\bibitem{erdogan2016improved}
H.~Erdogan, J.~R. Hershey, and et~al., ``Improved {MVDR} beamforming using
  single-channel mask prediction networks.'' in \emph{Interspeech}, 2016.

\bibitem{xu2019joint}
Y.~Xu, C.~Weng, L.~Hui, J.~Liu, M.~Yu, D.~Su, and D.~Yu, ``Joint training of
  complex ratio mask based beamformer and acoustic model for noise robust
  {ASR},'' in \emph{ICASSP}, 2019.

\bibitem{aswin2020icassp}
A.~S. Subramanian, C.~Weng, M.~Yu, S.-X. Zhang, Y.~Xu, S.~Watanabe, and D.~Yu,
  ``Far-field location guided target speech extraction using end-to-end speech
  recognition objectives,'' in \emph{ICASSP}, 2020.

\bibitem{wang2018mask}
Z.-Q. Wang and D.~Wang, ``Mask weighted {STFT} ratios for relative transfer
  function estimation and its application to robust {ASR},'' in \emph{ICASSP},
  2018.

\bibitem{boeddeker2018exploring}
C.~Boeddeker, H.~Erdogan, T.~Yoshioka, and R.~Haeb-Umbach, ``Exploring
  practical aspects of neural mask-based beamforming for far-field speech
  recognition,'' in \emph{ICASSP}, 2018.

\bibitem{xiao2017time}
X.~Xiao, S.~Zhao, D.~L. Jones, E.~S. Chng, and H.~Li, ``On time-frequency mask
  estimation for {MVDR} beamforming with application in robust speech
  recognition,'' in \emph{ICASSP}, 2017.

\bibitem{schasse2014estimation}
A.~Schasse and R.~Martin, ``Estimation of subband speech correlations for noise
  reduction via {MVDR} processing,'' \emph{IEEE/ACM Transactions on Audio,
  Speech, and Language Processing}, vol.~22, no.~9, pp. 1355--1365, 2014.

\bibitem{huang2011multi}
Y.~A. Huang and J.~Benesty, ``A multi-frame approach to the frequency-domain
  single-channel noise reduction problem,'' \emph{IEEE Transactions on Audio,
  Speech, and Language Processing}, vol.~20, no.~4, pp. 1256--1269, 2011.

\bibitem{tammen2019dnn}
M.~Tammen, D.~Fischer, and S.~Doclo, ``{DNN}-based multi-frame {MVDR} filtering
  for single-microphone speech enhancement,'' \emph{arXiv preprint
  arXiv:1905.08492}, 2019.

\bibitem{benesty2011speech}
J.~Benesty, J.~Chen, and E.~A. Habets, \emph{Speech enhancement in the {STFT}
  domain}.\hskip 1em plus 0.5em minus 0.4em\relax Springer Science \& Business
  Media, 2011.

\bibitem{fischer2017sensitivity}
D.~Fischer and S.~Doclo, ``Sensitivity analysis of the multi-frame {MVDR}
  filter for single-microphone speech enhancement,'' in \emph{EUSIPCO}, 2017,
  pp. 603--607.

\bibitem{habets2013two}
E.~A. Habets and J.~Benesty, ``A two-stage beamforming approach for noise
  reduction and dereverberation,'' \emph{IEEE Transactions on Audio, Speech,
  and Language Processing}, vol.~21, no.~5, pp. 945--958, 2013.

\bibitem{subramanian2019investigation}
A.~S. Subramanian, X.~Wang, S.~Watanabe, T.~Taniguchi, D.~Tran, and Y.~Fujita,
  ``An investigation of end-to-end multichannel speech recognition for
  reverberant and mismatch conditions,'' \emph{arXiv preprint
  arXiv:1904.09049}, 2019.

\bibitem{wang2014training}
Y.~Wang, A.~Narayanan, and D.~Wang, ``On training targets for supervised speech
  separation,'' \emph{IEEE/ACM transactions on audio, speech, and language
  processing}, vol.~22, no.~12, pp. 1849--1858, 2014.

\bibitem{yu2020audio}
J.~Yu, B.~Wu, and et~al., ``Audio-visual multi-channel recognition of
  overlapped speech,'' \emph{arXiv preprint arXiv:2005.08571}, 2020.

\bibitem{williamson2015complex}
D.~S. Williamson, Y.~Wang, and D.~Wang, ``Complex ratio masking for monaural
  speech separation,'' \emph{IEEE/ACM transactions on audio, speech, and
  language processing}, vol.~24, no.~3, pp. 483--492, 2015.

\bibitem{kinoshita2017neural}
K.~Kinoshita, M.~Delcroix, H.~Kwon, T.~Mori, and T.~Nakatani, ``Neural
  network-based spectrum estimation for online {WPE} dereverberation.'' in
  \emph{Interspeech}, 2017.

\bibitem{ephrat2018looking}
A.~Ephrat, I.~Mosseri, O.~Lang, T.~Dekel, K.~Wilson, A.~Hassidim, W.~T.
  Freeman, and M.~Rubinstein, ``Looking to listen at the cocktail party: A
  speaker-independent audio-visual model for speech separation,''
  \emph{SIGGRAPH}, 2018.

\bibitem{afouras2018conversation}
T.~Afouras, J.~S. Chung, and A.~Zisserman, ``The conversation: Deep
  audio-visual speech enhancement,'' \emph{Interspeech}, 2018.

\bibitem{morrone2019face}
G.~Morrone, S.~Bergamaschi, L.~Pasa, L.~Fadiga, V.~Tikhanoff, and L.~Badino,
  ``Face landmark-based speaker-independent audio-visual speech enhancement in
  multi-talker environments,'' in \emph{ICASSP}, 2019.

\bibitem{chung2016lip}
J.~S. Chung and A.~Zisserman, ``Lip reading in the wild,'' in \emph{Asian
  Conference on Computer Vision}, 2016.

\bibitem{stafylakis2017combining}
T.~Stafylakis and G.~Tzimiropoulos, ``Combining residual networks with {LSTM}s
  for lipreading,'' \emph{arXiv preprint arXiv:1703.04105}, 2017.

\bibitem{chen2018multi}
Z.~Chen, X.~Xiao, T.~Yoshioka, H.~Erdogan, J.~Li, and Y.~Gong, ``Multi-channel
  overlapped speech recognition with location guided speech extraction
  network,'' in \emph{2018 IEEE {SLT}}, 2018.

\bibitem{habets2006room}
E.~A. Habets, ``Room impulse response generator,'' \emph{Technische
  Universiteit Eindhoven, Tech. Rep}, vol.~2, no. 2.4, 2006.

\bibitem{yitu}
\url{https://speech.yitutech.com}.

\end{thebibliography}


\end{document}